\documentclass[12pt,letterpaper,oneside,onecolumn]{article}
\usepackage[]{fontenc}
\usepackage[dvips]{epsfig}

\makeatletter
\def\LyX{L\kern-.1667em\lower.25em\hbox{Y}\kern-.125emX\spacefactor1000}%
\newcommand{\lyxtitle}[1] {\thispagestyle{empty}
\global\@topnum\z@
\section*{\LARGE \centering \sffamily \bfseries \protect#1 }
}

\makeatother

\begin{document}

\baselineskip .3in \begin{titlepage}

{\centering \bfseries \large A Self-organising Model of Market
with Single Commodity\par}

{\centering \vskip .2in \par}

{\centering \bfseries Anirban Chakraborti \mdseries \( ^{(1)} \), \bfseries Srutarshi
Pradhan \mdseries \( ^{(2)} \) and \bfseries Bikas K. Chakrabarti \mdseries \( ^{(3)} \)\\
 \em Saha Institute of Nuclear Physics\em ,\\
 \em 1/AF Bidhan Nagar, Calcutta 700 064, India.\em \\
\par}

\noindent \vskip .3in 

\noindent \bfseries Abstract\mdseries 

\noindent We have studied here the self-organising features of the dynamics
of a model market, where the agents `trade' for a single commodity
with their money. The model market consists of fixed numbers of economic
agents, money supply and commodity. We demonstrate that the model,
apart from showing a self-organising behaviour, indicates a crucial
role for the money supply in the market and also its self-organising
behaviour is seen to be significantly affected when the money supply
becomes less than the optimum. We also observed that this optimal
money supply level of the market depends on the amount of `frustration'
or scarcity in the commodity market.\\
\vskip .3in 

 \end{titlepage}

\noindent \bfseries I. Introduction and the model\mdseries 

\noindent An economic market is perhaps the most commonly encountered self-organising
system or network, whose dynamics profoundly affects us all. Its dynamics
is no doubt very intriguing. In fact Adam Smith in 1776 first made
the formal notice of a self-organising aspect, which he called the
`invisible hand' effect, of the market consisting of selfish agents
[1, 2]. Although it still remains illusive to demonstrate that a pure
competitive market, consisting of agents pursuing pure self-interest,
can self-organise or reach (dynamic) equilibrium, the mainstream economists
seem to consider it to be true in principle (a matter of faith?) [2].
In various statistical physics models of interacting systems or networks,
such self-organisation has indeed been demonstrated to emerge in the
global aspects of the system which consists of a large number of simple
dynamical elements having local (in time and space) interactions and
dynamics [3]. It was also noted by the economists long time back that
this dynamics, which takes the system to equilibrium, is greatly facilitated
by `paper money' (rather than the direct commodity exchanges as in
barter economy) which does not have any value of its own, but rather
can be considered as a good `lubricant' in the econodynamics [2].
Consistent with this analogy, it was also seen that when the (paper)
money supply gets changed, it does not just scale up (for increased
money supply) or down (for decreased money supply) the commodity prices,
the (self-organising) dynamics towards equilibrium gets seriously
affected [4]. 

We have studied here the self-organising features of the dynamics
of a model market, where the agents `trade' for a single commodity
with their money. We demonstrate that the model, apart from having
a self-organising behaviour, has got a crucial role for the money
supply in the market and that its self-organising behaviour is significantly
affected when the money supply becomes less than the optimum. We also
observed that this optimal money supply level of the market depends
on the amount of `frustration' or scarcity in the commodity market.
In our model, each agent having commodity less than the `subsistence'
level trades with any other having more than the `subsistence' level,
in exchange of its money. Specifically, we consider a closed economic
system consisting of \( N \) economic agents, each economic agent \( i \) have
at any time money \( m_{i} \) and commodity \( q_{i} \), such that (\( \sum ^{N}_{i=1}m_{i}=M \) and \( \sum ^{N}_{i=1}q_{i}=Q \)), where \( N \),
\( M \) and \( Q \) are fixed. The `subsistence' commodity level for each agent
is \( q_{0} \). Hence at any time an agent having \( q_{i}<q_{0} \) will attempt to trade, utilising
its money \( m_{i} \) at that time, with agents having commodity more than \( q_{0} \),
and will purchase to make its commodity level \( q_{0} \) (and no-further),
if its money permits. The agents with \( q_{i}>q_{0} \) will sell-off the excess amount
to such `hungry' agents having \( q_{i}<q_{0} \), and will attempt to maximize its
wealth (money). This dynamics is local in time (`daily') and it stops
eventually when no further trade is possible satisfying the above
criteria. We introduce an `annual' or long-time dynamics when some
random fluctuations in all the agents' money and commodity occur.
Annually, each agent gets a minor reshuffle of the money and its commodity
(e.g., perhaps due to the noise in the stock market and in the harvest
due to the changes in the weather respectively). This (short and long
time) combined dynamics is similar to that of the `sand-pile' models
studied extensively in recent times [3]. The price of the commodity
does not change in our model with the money supply \( M \) in the market;
it remains fixed here (at unity). We look for the steady state features
of this market; in particular, the distributions \( P(m) \) and \( P(q) \) of the money
and commodity respectively among the agents. We have investigated
how many agents \( P(q_{0}) \) can satisfy their basic needs through this dynamics,
i.e., can reach the subsistence level \( q_{0} \), as function of the money
supply \( M \) for both unfrustrated (\( g<1 \)) (where \( g=q_{0}/<q> \) and \( <q> \) \( =Q/N \) is the average
commodity in the market) and the frustrated (\( g>1 \)) cases of the commodity
market. We observed that an optimum amount \( M \)\( _{0} \) of money supply is required
for evolving the market towards the maximum possible value of \( P(q_{0}) \). This
optimum value of money \( M \)\( _{0} \) is observed to decrease with increase in
\( g \) in the market. 

\vskip .2in 

\noindent \bfseries II. Unlimited money supply and limited supply of commodity
: \mdseries 

Here, we consider the money supply \( M \) in the market to be infinitely
large and it therefore drops out from any consideration. The dynamics
is then entirely governed by the commodity distribution among agents:
for agents with \( q_{i}<q_{0} \), the attempt will be to find another trade partner
having \( q_{i}>q_{0} \); and having found such partners, through random search in
the market, trades occur for mutual benefit (for the selling agent
we still consider the extra money from trade to be important). The
system thus evolves towards its steady state, as the fixed-point feature
of the short-time or daily dynamics gets affected by the random noise
reshuffling in the commodity of each agent. This reshuffling essentially
induces Gibbs-like distribution [5, 6]. The trade dynamics is clearly
motivated or `directed'. We look for the combined effect on the steady
state distribution of commodity \( P(q) \), which is independent of the initial
commodity distribution among the agents.

For the unfrustrated case (\( g<1 \)), where \em all \em the agents can be
satisfied, the typical distributions \( P(q) \) are shown in Fig. 1 for different
values of \( g \). We see that the \( P(q) \) is Gibbs-like (\( P(q) \) \( =A\exp (-q/<q>) \) and \( A=1-g \)), for \( q>q_{0} \), while
\( P(q_{0})=g \) (as shown in the inset). One can easily explain these observations
using the fact that the cumulative effect of the long-time randomization
gives Gibbs distribution (\( \exp (-q/<q>) \) for all \( q \)). We then estimate the final
steady state distribution \( P(q) \) from the additional effect of the short-time
dynamics on this (long-time dynamics induced) Gibbs distribution.
All the agents with \( q<q_{0} \) manage here to acquire \( q_{0} \)-level of commodity
(as \( g<1 \) and everybody has enough money to purchase the required amount).
Their number is then given by \( N_{-}=\int _{0}^{q_{0}}\exp (-q)dq \). They require the total amount of
commodity \( q_{0}N_{-} \). The amount of commodity already available with them is
given by \( Q_{-}=\int _{0}^{q_{0}}q\exp (-q)dq \). The excess amount required \( Q_{demand}=q_{0}N_{-}-Q_{-} \) has to come from the agents
having \( q>q_{0} \). The average of the excess amount of commodity of the agents
who are above the \( q_{0} \)-line is given by \( <q_{excess}>=(1-Q_{-}-(1-N_{-})q_{0})/(1-N_{-}) \). The number of agents who supply
the \( Q_{demand} \) amount is given by \( N_{+}=Q_{demand}/<q_{excess}> \). This gives \( P(q_{0})=N_{-}+N_{+} \)\( =g \). We can easily determine
the prefactor of the final steady state distribution \( P(q) \) for \( q>q_{0} \), \( A=1-g \) from
the conservation of total number of agents and total commodity.

For the frustrated case (\( g>1 \)), the results are shown in Fig. 2. A similar
calculation for \( P(q_{0}) \) is done as follows: \( N_{+}=\int _{q_{0}}^{\infty }\exp (-q)dq \) is the number of people above
the \( q_{0} \)-line who will sell off their excess amount of commodity to come
to \( q_{0} \)-level, \( Q_{+}=\int _{q_{0}}^{\infty }q\exp (-q)dq \) is the commodity of the agents above the \( q_{0} \)-level. Then
the supplied amount of commodity to the agents below the \( q_{0} \)-line is
\( Q_{supply}=Q_{+}-q_{0}N_{+} \). The average of the deficit commodity, \( <q_{deficit}>=((1-N_{+})q_{0}-1+Q_{+})/(1-N_{+}) \). Hence, the number of agents
who will reach \( q_{0} \)-level from below is \( N_{-}=Q_{supply}/<q_{deficit}> \) , so that \( P(q_{0})=N_{+}+N_{-}= \)\( g\exp (-g)/(g-1+\exp (-g)) \). A comparison
of this estimate for \( P(q_{0}) \) with \( g \) is also shown in the inset of Fig. 2.
It may be mentioned that in absence of the strict Gibbs distribution
for \( P(q) \) ( \( q<q_{0} \)), the above expression for \( P(q_{0}) \) is somewhat approximate.

\vskip .2in 

\noindent \bfseries III. Limited money supply and limited supply of commodity
:\mdseries 

When the money supply is limited, the self-organising behaviour is
significantly affected and the fraction of agents who can secure \( q_{0} \)
amount of commodity for themselves \( P(q_{0}) \), does not always reach its maximum
possible value (as suggested by the amount of commodity available
in the market). For all values of \( g \), as we increase the money supply
in the market \( M \), \( P(q_{0}) \) increases and then after a certain amount \( M_{0} \), it
saturates. In Fig. 3., we have shown how the quantity \( P(q_{0}) \) varies with
\( M \) for different values of \( g \) (for \( g>1 \) only, as we are more interested
in the frustrated case). We define \( M_{0} \) to be the optimum amount of money
supply needed for the smooth functioning of the market. We also observed
that this optimal money supply level of the market depends on the
amount of frustration \( g \) in the commodity market. In the inset, the
variation of \( M_{0} \) with \( g \) is shown for the frustrated case (\( g>1 \)) only.

\vskip .2in 

\noindent \bfseries IV. Summary and Concluding remarks\mdseries 

We have studied here the steady state distributions \( P(m) \) and \( P(q) \) of money
and commodity in a model consisting of fixed number of agents \( N \), total
commodity \( Q \) and total money \( M \) in the market. Only one commodity is
considered for trade and its price is taken to be fixed (at unity)
and it does not change with the money supply in the market. The subsistence
commodity level \( q_{0} \) of all the agents are the same and each of them
would like to purchase the deficit amount (\( q_{0}-q_{i} \)) from the agents having
\( q_{i}>q_{0} \), in exchange of its own money. Apart from the basic urge to reach
the subsistence level (if originally below it), all the agents would
like to maximize their wealth or money. The second instinct allows
the agents with excess commodity (over \( q_{0} \)) to find hungry partners
and to sell-off the excess. The dynamics considered here is the local
or short time (or daily) dynamics. Additionally, we consider a long
time (or yearly) dynamics, which reshuffles mildly but randomly the
amounts of money and commodity of each agent (due to, say, the fluctuations
in the stock market and in the harvest, etc.). The resulting distributions
follow from the successive applications of the local directed dynamics,
followed by a randomization in the quantities. We consider both the
cases: unfrustrated case (\( g<1 \)) where in principle every agent can be
satisfied, and frustrated case (\( g>1 \)) where not everyone can be satisfied.
We concentrate mainly on the quantity \( P(q_{0}) \) which gives the steady density
of agents in the market who can satisfy the basic requirement of commodity.
As is obvious, this quantity is most significant in the frustrated
cases (\( g>1 \)) where there is not enough commodity in the market to satisfy
the basic requirements (\( q_{0} \)) for everyone. We study how this quantity
grows or the distribution of commodity among the agents is facilitated,
with the supply of money \( M \) in the market.

We see that \( P(q_{0})=g \) for \( g<1 \) and \( P(q_{0})\simeq g\exp (-g)/(g-1+\exp (-g)) \) for \( g>1 \), where the money supply \( M \) is much
greater than \( M_{0} \), the optimal money required in the market. We see that
these expressions for the \( P(q_{0}) \) for \( M>M_{0} \) can be easily explained (see section
II) assuming that the resulting distribution is Gibbs like for \( q>q_{0} \) (for
\( g<1 \)) and a mean-field picture (for \( g>1 \)). 

\noindent \newpage

\noindent \bfseries References\mdseries 

\noindent \vskip .1in 

\noindent \em e-mail addresses \em :

\noindent \( ^{(1)} \) anirban@cmp.saha.ernet.in

\noindent \( ^{(2)} \) spradhan@cmp.saha.ernet.in 

\noindent \( ^{(3)} \) bikas@cmp.saha.ernet.in

\noindent \vskip .2 in

\noindent 1. A. Smith, \em An Inquiry into the Nature and Causes of The Wealth
of Nations\em , Strahan and Caddell, London (1776).

\noindent 2. See e.g., P. A. Samuelson, \em Economics\em , \( 16 \)th Edition, McGraw-Hill
Inc., Auckland, 29-32 (1998). 

\noindent 3. See e.g., P. Bak, \em How Nature Works: The Science of Self-Organised
Criticality\em , Springer, New York (1996).

\noindent 4. J. M. Keynes, \em General Theory of Employment, Interest and Money\em ,
Royal Economic Society, Macmillan Press, London (1973).

\noindent 5. A. Dragulescu and V. M. Yakovenko, \em Eur. Phys. J. B\em , \bfseries 17\mdseries ,
723-729 (2000).

\noindent 6. A. Chakraborti and B. K. Chakrabarti, \em Eur. Phys. J. B\em ,
\bfseries 17\mdseries , 167-170 (2000).

\noindent \newpage

\noindent \bfseries Figure captions\mdseries \\
 \vskip .1 in 

\noindent \bfseries Fig. 1 \mdseries : The distributions of commodity
\( P(q) \) for different values of \( g \) for \( N=1000 \), \( Q=1 \), \( M=100 \) (\( M>M_{0}(g) \)), for the unfrustrated
case (\( g<1 \)). The steady-state distribution of commodity \( P(q) \) is Gibbs-like:
\( P(q) \) \( =A\exp (-q/<q>) \) with \( A=1-g \), for \( q>q_{0} \). The inset shows the linear variation of \( P(q_{0}) \) with
\( g \) \( (P(q_{0})=g \)).

\noindent \bfseries Fig. 2 \mdseries : The distributions of commodity
\( P(q) \) for different values of \( g \) for \( N=1000 \), \( Q=1 \), \( M=100 \) (\( M>M_{0}(g) \)), for the frustrated case
(\( g>1 \)). The variation of \( P(q_{0}) \) with \( g \) is shown in the inset where the theoretical
estimate (\( P(q_{})=g\exp (-g)/(g-1+\exp (-g)) \)) is also indicated by the solid line.

\noindent \bfseries Fig. 3 \mdseries : The variation of quantity \( P(q_{0}) \) with
\( M \) for different values of \( g \) in the frustrated cases (\( g>1 \)). The inset
shows the variation of \( M_{0} \) with \( g \).\\

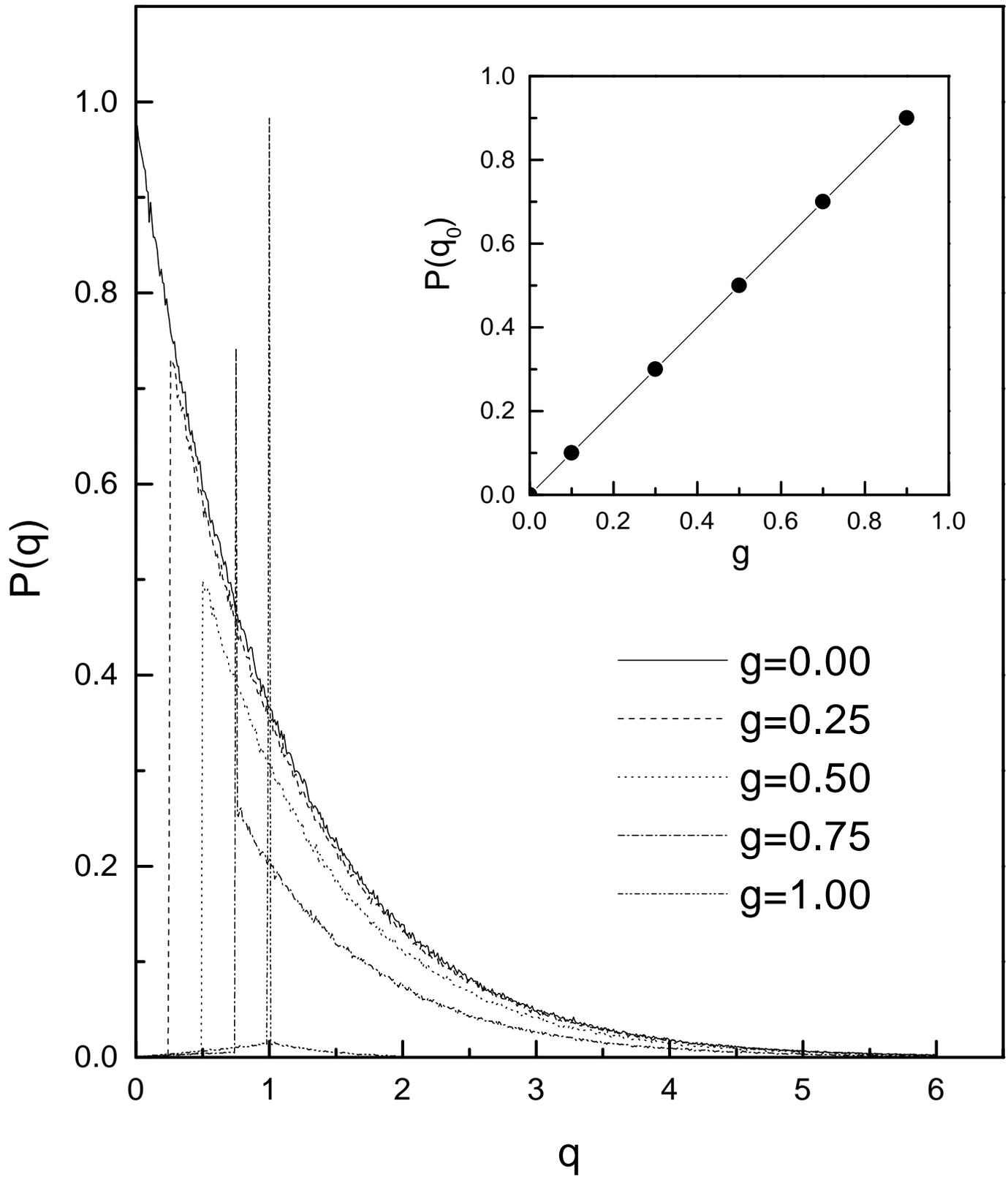
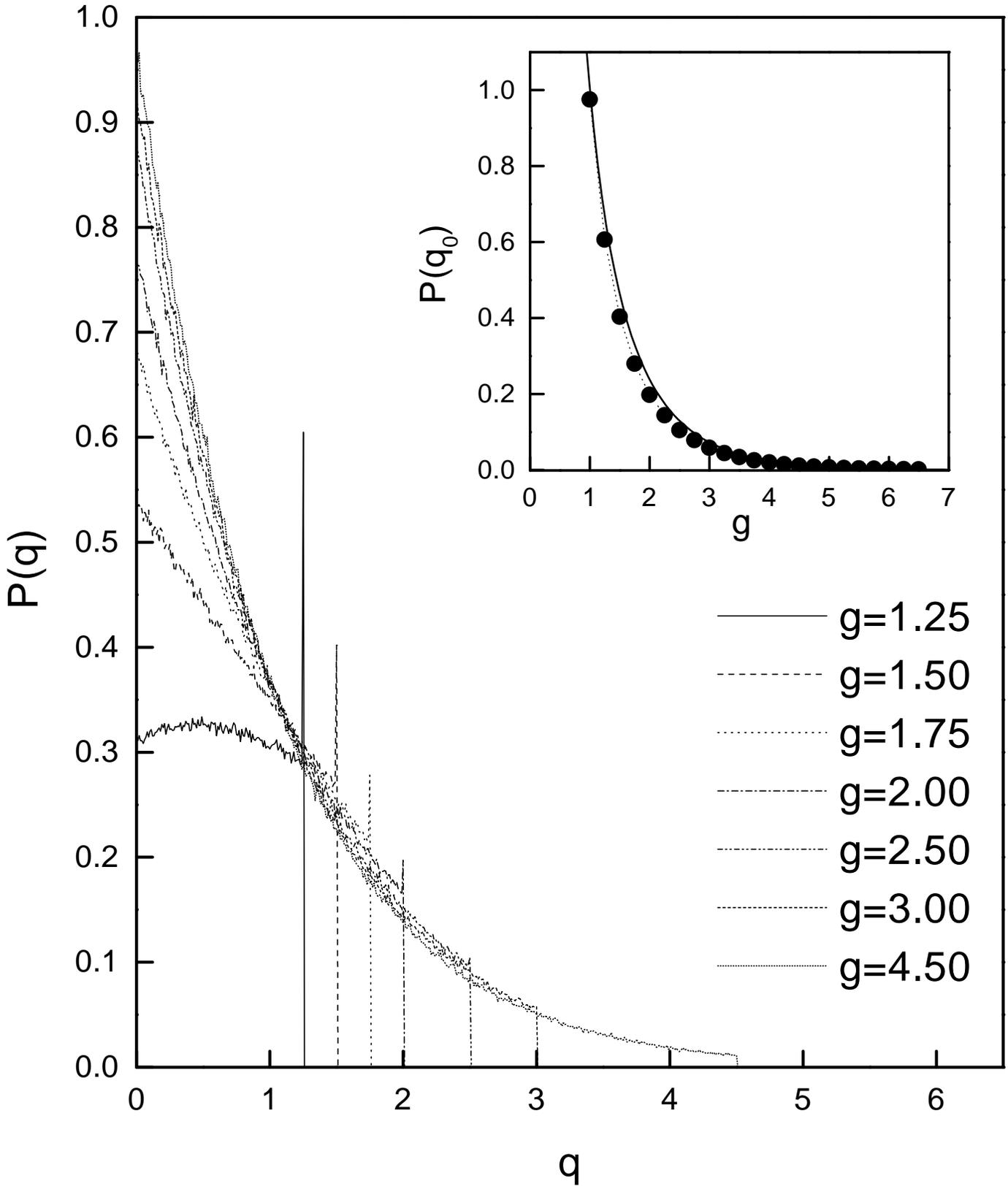
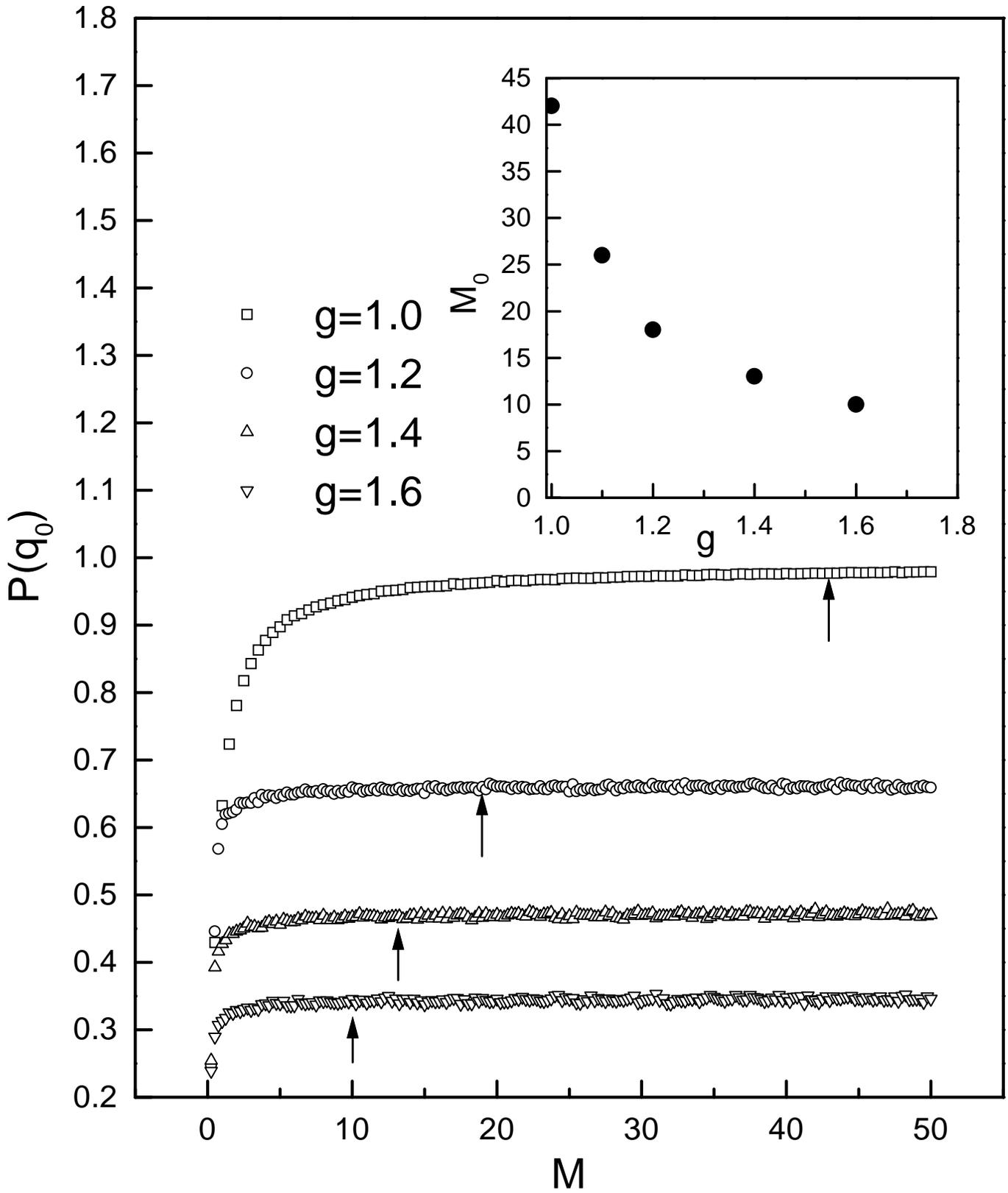
\end{document}